\documentclass[12pt]{article}
\usepackage{epsf}
\newcommand{\be}{\begin{equation}}
\newcommand{\ee}{\end{equation}}
\newcommand{\bea}{\begin{eqnarray}}
\newcommand{\eea}{\end{eqnarray}}
\newcommand{\ba}{\begin{array}}
\newcommand{\ea}{\end{array}}

\title{ {\bf
Estimate of the long-distance contribution through $b\rightarrow s \psi$
to the $B_{s}\rightarrow\gamma\gamma$ decay rate}}
\author{\vspace{1cm}\\
        \vspace{5mm}\\
        {\bf G. Hiller}\thanks{E-mail address: ghiller@x4u2.desy.de} , \\
         Deutsches Elektronen-Synchrotron DESY, Hamburg \\
        \vspace{5mm}\\
        {\bf E. O. Iltan}
        \thanks{E-mail address:
        eiltan@heraklit.physics.metu.edu.tr}
 \\
        Physics Department, Middle East Technical University \\
        Ankara, Turkey\\}

\date{}
\begin{document}
\setlength{\baselineskip}{24pt}
 \maketitle

 \begin{abstract}
 \baselineskip  .7cm
 The $B_{s}\rightarrow \phi\psi$ decay is modeled through the inclusive 
$b\rightarrow s\psi$ decay. Using the Vector Meson Dominance model,
the amplitude for the chain process 
$B_{s}\rightarrow\phi\psi\rightarrow\phi\gamma\rightarrow\gamma\gamma$
is estimated and it is found to be at most $4 \%$ of the corresponding
amplitude from the $O_7$ type LD contribution.
The intermediate amplitude for the process
$B_{s}\rightarrow\phi\gamma$ is compared with the
corresponding one obtained by a different approach based on the interaction 
of the virtual charm quark
loop with soft gluons \cite{Ruckl}.
Both amplitudes are found to agree within $10 \%$.
Further the influence on the branching ratio
$B(B_{s}\rightarrow\gamma\gamma)_{SD+LD_{O7}}$ 
from inclusive $b\rightarrow s \psi$
is estimated as less than $1 \%$.
 \end{abstract}
\thispagestyle{empty}
\newpage
\setcounter{page}{1}

\section{Introduction}
Rare B-meson decays are one of the main research fields in recent 
particle physics.
Since they occur only at loop level in common flavour changing neutral 
current (FCNC)- forbidden models, 
the theoretical and experimental investigations provide precise tests 
of the Standard Model (SM) and possible new physics beyond.
An interesting candidate for rare decays is
$B_{s}\rightarrow \gamma \gamma$, which has a rich final state
since the two photon system can be in a CP-odd or a CP-even state.
In the literature, $B_{s}\rightarrow \gamma \gamma$ 
has been studied in the lowest order in the framework of the 
constituent quark model \cite{yaosimmaaliev}
and found to 
consist in the effective Hamiltonian theory \cite{effham}, which is obtained by
integrating out heavy particles
(top quark and $W$ boson in the SM) 
from the full theory, of
contributions from $O_7$ and four-quark operators. 
The calculation for $B_{s}\rightarrow \gamma \gamma$ decay 
with leading logarithmic QCD corrections 
including $O_{7}$ type long distance (LD) effects
and in an 
approach based on heavy quark effective theory to model the bound state 
was recently done in \cite{gudi}.

In the present work we continue and estimate the additional LD effect 
due to the dominant four-quark operators $O_1$ and $O_{2}$ 
(see eq.~(\ref{O12})) through the 
$B_{s}\rightarrow \phi\psi\rightarrow\phi\gamma\rightarrow\gamma \gamma$
chain decay. 
We use at quark level $b\rightarrow s \psi$ followed by the 
$b\rightarrow s\gamma$ decay \cite{desphande} and we pass to the 
hadronic level using the transition form factor $F_{1}(0)$
from the amplitude ${\cal{A}}(B_{s}\rightarrow\phi)$
\cite{gudi},\cite{alibraun}.
For both the conversions $\psi\rightarrow\gamma$ and 
$\phi\rightarrow\gamma$ we impose the Vector Meson Dominance (VMD) model 
\cite{desphande}. The conversion $\psi\rightarrow\gamma$ needs further
manipulation because of the strong contribution of the longitudinal part
of the $\psi$ meson. We extract the transverse part using the Golowich-Pakvasa
procedure \cite{desphande}, \cite{pakvasa}.
Further we calculate the $O_{1,\, 2}$ type LD effect to the $B_{s}\rightarrow
\phi\gamma$ decay using the method given in \cite{Ruckl}, namely, by taking
into account the virtual c-quark loop instead of the hadronization of the
$\bar{c}c$ pair. This procedure was originally applied to estimate the 
LD effect in $B \to K^{\ast} \gamma$ decay and uses operator product 
expansion and QCD sum rule techniques. We compare the amplitudes 
obtained by these two 
different methods and show that they are in good agreement within the errors 
of the calculation.
Finally we estimate the influence of the $O_{1,\, 2}$ type LD contribution
on the branching ratio for $B_s \to \gamma \gamma$ decay.
 
The paper is organized as follows:
In section 2, we calculate the $O_{1,2}$ type LD contribution to 
$B_s\rightarrow \gamma\gamma$ due to the chain process 
$B_{s}\rightarrow\phi\psi \rightarrow\phi\gamma\rightarrow\gamma\gamma$ 
using the Gordon
decomposition and further the VMD model. 
We compare our amplitude for the decay 
$B_{s}\rightarrow\phi\psi \rightarrow\phi\gamma$ with the $O_{1,\,2}$ type 
one for $B_{s}\rightarrow\phi\gamma$, calculated by 
virtual charm loop-soft gluon interaction \cite{Ruckl}.   
Section 3 contains a discussion of the
branching ratio 
${\cal{B}}(B_{s}\rightarrow\gamma\gamma)_{SD+LD_{O_{7}}+\bar{LD}_{O_{2}}}$
where the subscripts 
$SD, \, LD_{O_7}, \,  \bar{LD}_{O_{2}}$ denote
QCD short distance (SD) contributions and LD ones due to
$B_s \rightarrow\phi\gamma \rightarrow\gamma\gamma$  and 
$B_{s}\rightarrow\phi\psi \rightarrow\phi\gamma\rightarrow\gamma\gamma$ 
decays, respectively.

\section{The chain process
$B_{s}\rightarrow\phi\psi\rightarrow\phi\gamma\rightarrow\gamma\gamma$}

\begin{figure}[htb]
\vskip -0.6truein
\centering
\epsfysize=7in
\leavevmode\epsffile{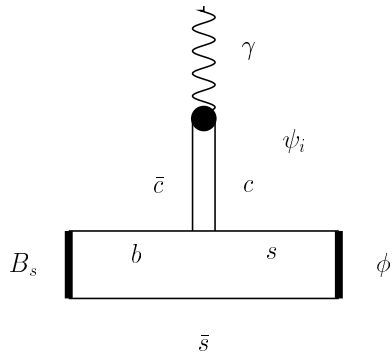}
\vskip -4.2truein
\caption[]{ The diagram contributing to 
$B_s \rightarrow \phi \psi \rightarrow \phi \gamma $. }
\label{fig:vmd}
\end{figure}

We first consider the additional contribution to $b\rightarrow s\gamma$ 
from $b\rightarrow s\psi_{i} \rightarrow s \gamma$, 
where $\psi_{i}$ are all $\bar{c}c$
$J=1$ bound states, see fig.~\ref{fig:vmd}.
The relevant part of the effective Hamiltonian describing this process
is given as
\begin{eqnarray}
{\cal{H}}_{eff}=4 \frac{G_{F}}{\sqrt{2}} V^{*}_{cs} V_{cb}
(C_{1}(\mu) O_{1}(\mu)+C_{2}(\mu) O_{2}(\mu)),
\label{effH2}
\end{eqnarray}
with the dominant four-Fermi operators 
\begin{eqnarray}
O_{1}&=&\bar{s}_{\alpha}\gamma_{\mu}\frac{1-\gamma_{5}}{2}c_{\beta}\,\,\bar{c}_{\beta}
\gamma_{\mu}\frac{1-\gamma_{5}}{2}b_{\alpha}\nonumber \, \, ,\\
O_{2}&=&\bar{s}\gamma_{\mu}\frac{1-\gamma_{5}}{2}c\,\,\bar{c}\gamma_{\mu}
\frac{1-\gamma_{5}}{2}b \, \, .
\label{O12}
\end{eqnarray}
Here $\alpha,\beta$ are $SU(3)$ colour indices
and $ V^{(*)}_{ij}$
are the relevant elements of the quark mixing matrix.
The initial values of the corresponding Wilson coefficients are
$C_{1}(m_W)=0$ and $C_{2}(m_W)=1$.
To include leading logarithmic QCD corrections we evaluate
$C_{1,2}(\mu)$ at the relevant scale, $\mu \approx m_b$ for $B$-decays,
and this takes into account short distance effects from single gluon exchange.
The analytical expressions can be found in \cite{effham}. 

Further we have used the unitarity of the CKM matrix
$V_{cs}^{*}V_{cb}=-V_{ts}^{*}V_{tb}-V_{us}^{*}V_{ub}$ and have neglected 
the contribution due to an internal u-quark, since 
$V_{us}^{*}V_{ub}<< V_{ts}^{*}V_{tb}=\lambda_t$.

Using factorization, we obtain the inclusive decay amplitude 
for the process $b\rightarrow s\psi$ \cite{desphande} as
\begin{eqnarray}
{\cal{A}}(b\rightarrow s\,\psi(k_{1},\epsilon^{\psi}))=-i
C f_{\psi}(m^{2}_{\psi}) m_{\psi} 
\bar{s}\gamma^{\mu}(1-\gamma_{5})b\, \epsilon_{\mu}^{\psi} \, \, .
\label{A1}
\end{eqnarray}
Here 
\begin{eqnarray}
C=-\frac{G_{F}}{\sqrt{2}} \lambda_t a_2(\mu)
\label{const1}
\end{eqnarray}
with, assuming naive factorization,
\begin{eqnarray}
a_2(\mu)=C_{1}(\mu)+\frac{C_{2}(\mu)}{N_C} \, \, ,
\label{a2}
\end{eqnarray}
where $N_C=3$ in colour $SU(3)$ and
$k_1, \epsilon^{\psi}$ are the momentum and the polarization vector of the 
$\psi$, respectively.
In eq.~(\ref{A1}) we used the matrix element 
\begin{eqnarray}
<0|\bar{c}\gamma_{\mu} c|\psi(k_{1},\epsilon^{\psi})>=
f_{\psi} (m^{2}_{\psi}) m_{\psi}
\epsilon_{\mu}^{\psi} \, \, .
\label{psim}
\end{eqnarray}

At this stage there is a critical remark about factorization 
in order, concerning the value of $a_2(\mu)$ used.
The decay under consideration is a class II decay following the 
classification of \cite{BSW}. 
In general eq.~(\ref{a2}) is written as 
\begin{eqnarray}
a_2^{eff}=(C_{1}(\mu)+\frac{C_{2}(\mu)}{N_C}) 
\left[ 1+\epsilon_{1}(\mu) \right]+ 
C_{2}(\mu) \epsilon_{2}(\mu) \, \, ,
\label{a2eff}
\end{eqnarray}
where $\epsilon_{1}(\mu)$ and $\epsilon_{2}(\mu)$ parametrize the
non-factorizable
contributions to the hadronic matrix elements.
$a_{2}^{eff}$ takes into account all contributions of the matrix elements
in contrast to $a_{2}(\mu)$, which assumes naive factorization 
$\epsilon_1(\mu)=\epsilon_2(\mu)=0$.
Especially $\epsilon_{2}(\mu)$, which is the colour octet piece, has sizable
contributions to naive factorization in class II decays \cite{neubert}. 
Furthermore, the 
additional problem is not to know the correct factorization scale.
In order to include the non-factorizable corrections 
we use in the definition of $C$ in eq.~(\ref{const1}) 
the effective coefficient 
$a_{2}^{eff}$, which is determined experimentally
from the world average branching ratio of
$\bar{B} \to \bar{K}^{(\ast)} \psi $ as \cite{neubert} 
\begin{eqnarray}
a_2^{eff}=0.21 \, \, .
\label{a2eff2}
\end{eqnarray}
This choice restores the correct scale and is $\mu$ independent.
Writing the Ansatz
\begin{eqnarray}
a_{2}^{eff}=C_{1}(m_{b})+\xi C_{2}(m_{b}) \, \, ,
\label{a2xi}
\end{eqnarray}
it follows that $\xi\approx 0.41$ with 
$C_{1}(m_b)=-0.25$ and $C_{2}(m_b)=1.11$ for the input values given in 
Table~\ref{parameters}. For comparison, naive factorization would give 
$a_2(m_b)=0.12$. 
\begin{table}[h]
        \begin{center}
        \begin{tabular}{|l|l|}
        \hline
        \multicolumn{1}{|c|}{Parameter} & 
                \multicolumn{1}{|c|}{Value}     \\
        \hline \hline
        $m_b$                   & $4.8$ (GeV) \\
         $m_c$                   & $1.4$ (GeV) \\
        $\alpha_{em}^{-1}$      & 129           \\
        $\lambda_t$            & 0.04 \\
        $m_{t}$             & $175$ (GeV) \\
        $m_{W}$             & $80.26$ (GeV) \\
        $m_{Z}$             & $91.19$ (GeV) \\
        $\Lambda^{(5)}_{QCD}$             & $0.214$ (GeV) \\
        $\alpha_{s}(m_Z)$             & $0.117$  \\
        \hline
        \end{tabular}
        \end{center}
\caption{Values of the input parameters used in the numerical
          calculations.}
\label{parameters}
\end{table}

Now our aim is to replace the $\psi$ meson with the photon $\gamma$ and
to construct a gauge invariant amplitude. 
This can
be done by killing the longitudinal component of the meson $\psi$ so, that 
$\epsilon_{\mu}^{\psi}$ is changed to the
polarization vector $\epsilon_{\mu}^{\gamma}$ of the photon $\gamma$. For this
we use the Golowich-Pakvasa \cite{desphande},\cite{pakvasa} procedure with the help of the 
Gordon identity, namely
$\gamma_{\mu}\gamma_{\alpha}=g_{\mu\alpha}-i\sigma_{\mu\alpha}$.
We start with the vertex $\bar{s}\gamma_{\mu}(1-\gamma_{5})b$ and using the 
equation of motion ${\not{p}} b=m_b b$ and momentum conservation $p=p'+k_1$, 
we get
\begin{eqnarray}
\bar{s}\gamma_{\mu}(1-\gamma_{5})b=\frac{1}{m_{b}}\{\bar{s}\gamma_{\mu}\not{p'}
(1+\gamma_{5})b + \bar{s}\gamma_{\mu}\not{k_{1}} (1+\gamma_{5})b \} \, \, ,
\label{vert1}
\end{eqnarray}
where $p,p'$ are the momenta of the $b$ and $s$ quark, respectively.
We neglect the first term in eq.~(\ref{vert1}) since 
$\frac{m_{s}}{m_b} \ll 1$ and $p'^{\mu} \epsilon^{T}_{\mu}=0$, which follows 
from $\epsilon_{\mu}^{T} p^{\mu}=0$ 
in the rest frame of the $b$ quark and the transversality condition
$\epsilon_{\mu}^{T} k_{1}^{\mu}=0$, 
where $\epsilon_{\mu}^{T}$ is the transversal polarization 
vector of the $\psi$ meson \cite{desphande}.
The second term can be written as
\begin{eqnarray}
\frac{1}{m_{b}}\bar{s}\gamma_{\mu}\not{k}_{1} (1+\gamma_{5}) b=
\frac{1}{m_{b}} \{ \bar{s}(1+\gamma_{5}) k_{1 \mu} b 
-i \bar{s}\sigma_{\mu\alpha}k_{1}^{\alpha}(1+\gamma_{5})b \} \, \, .
\label{vert2}
\end{eqnarray}
Only the $\sigma_{\mu\alpha}$ term in eq.~(\ref{vert2}) couples to 
the transversal component of the $\psi$ and we obtain the corresponding 
amplitude as
\begin{eqnarray}
{\cal{A}}(b\rightarrow s \psi^{T})=
-2 C f_{\psi}(m^{2}_{\psi}) \frac{m_{\psi}}{m_{b}}
\bar{s}\sigma_{\mu\alpha} k_{1}^{\alpha}\, R \, b \epsilon_{\mu}^{T},
\label{A2}
\end{eqnarray}
where $R=\frac{1+\gamma_{5}}{2}$ denotes the chiral right projection.
Note that the coupling structure is the same as due to a direct use of 
$O_7= \frac{e}{16 \pi^2} \bar{s} \sigma_{\mu \nu} m_b R b F^{\mu \nu} $ 
\cite{effham}
with the photon field strength tensor $ F^{\mu \nu}$ and $m_s=0$.
For the $\psi^{T}\rightarrow\gamma$ conversion following the VMD mechanism 
we have
\begin{eqnarray}
<0|J_{\mu, el}|\psi^{T}(k_{1}, \epsilon^{T})0>=e Q_{c} f_{\psi}(0) m_{\psi}
\epsilon^{T}_{\mu} \, \, ,
\label{conv}
\end{eqnarray}
where $Q_{c}=2/3$ and $f_{\psi}(0)$ is the coupling 
at $k_1^2=0$, see eq.~(\ref{suppr}).
Using the intermediate propagator of the $\psi$ meson at $k_{1}^2=0$,
we get
\begin{eqnarray}
{\cal{A}}(b\rightarrow s \psi^{T}\rightarrow s\gamma )=
2 C f^{2}_{\psi}(0) 
\frac{e Q_{c}}{m_{b}}
\bar{s}\sigma_{\mu\alpha} k_{1}^{\alpha} \,R \, b \, \epsilon_{\mu}^{T}
\, \, .
\label{A3}
\end{eqnarray}
The expression for the amplitude eq.~(\ref{A3}) can be completed by 
summing over all 
$\bar{c}c$ resonant states $\psi(1S)$,$\psi(2S)$,$\psi(3770)$,$\psi(4040)$,
$\psi(4160)$ and $\psi(4415)$
\begin{eqnarray}
{\cal{A}}(b\rightarrow s \psi_{i}^{T}\rightarrow s\gamma )=
2 C \sum_{i} f^{2}_{\psi_{i}}(0) 
\frac{e Q_{c}}{m_{b}}
\bar{s}\sigma_{\mu\alpha} k_{1}^{\alpha} \,R \,b \epsilon_{\mu}^{T}
\label{A4} \, \, .
\end{eqnarray}
The various decay couplings $f_{\psi_{i}}=f_{\psi_{i}}(m^{2}_{\psi_{i}})$ are
calculated using 
\begin{eqnarray}
f^{2}_{\psi_{i}}=\Gamma(\psi_{i}\rightarrow e^{+}e^{-})\frac{3 m_{\psi_{i}}}
{Q_c^2 4 \pi\alpha^{2}_{em}}
\label{fpsi2}
\end{eqnarray}
and the measured widths from \cite{PDG}
and given in  Table~\ref{fpsi}. 
\begin{table}[h]
        \begin{center}
        \begin{tabular}{|l|l|}
        \hline
        \multicolumn{1}{|c|}{$\psi_i$} & 
                \multicolumn{1}{|c|}{$f_{\psi_{i}} [GeV]$}     \\
        \hline \hline
$f_{\psi(1S)}$& 0.405  \\
$f_{\psi(2S)}$& 0.282   \\
$f_{\psi(3770)}$& 0.099    \\
$f_{\psi(4040)}$& 0.175  \\
$f_{\psi(4160)}$& 0.180   \\
$f_{\psi(4415)}$& 0.145  \\
        \hline
        \end{tabular}
        \end{center}
\caption{Vector meson coupling constants used in the numerical
          calculations.}
\label{fpsi}
\end{table}

Here we need to extrapolate the coupling $f_{\psi_{i}}(k_1^{2}=m_{\psi_i}^{2})$
to $f_{\psi_i}(0)$. We used the suppression factor \cite{desphande} 
\begin{eqnarray}
\kappa=f^{2}_{\psi(1S)}(0)/f^{2}_{\psi(1S)}(m_{\psi}^{2})=0.12
\label{suppr}
\end{eqnarray}
obtained from data on the photoproduction of the $\psi$
and taking the factor $\kappa$ as universal for the other resonances.
\footnote{This is consistent with $\kappa=0.11$ \cite{terasaki} based on a
dispersion relation calculation.}
We now use eq.~(\ref{A4}) to find the matrix element of
$B_{s}\rightarrow\phi\gamma$ through the 
$b\rightarrow s\psi^{T}\rightarrow s \gamma$
transition at quark level. The matrix element \cite{alibraun} 
is given as
\begin{eqnarray}
<\phi(p')|\bar{s}\sigma_{\mu\alpha}\, R \, k_{1}^{\alpha} b|B_{s}(p)>&=&i
\epsilon_{\mu\nu\rho\sigma}\epsilon^{\phi
\nu}p^{\rho}p'^{\sigma}F_{1}(k_1^{2})\nonumber \\
&+& (\epsilon_{\mu}^{\phi} p.k_1-p_{\mu} k_1.\epsilon^{\phi}) G(k_1^{2}),
\label{alibraun}
\end{eqnarray}
and we get the amplitude 
\begin{eqnarray}
{\cal{A}}(B_{s}\rightarrow\phi\gamma)&=&
2 C \epsilon_{1}^{\mu} \epsilon^{\phi \nu} 
\sum_{i}\frac{f^{2}_{\psi_{i}}(0)}{m_{b}} e Q_{c} \{ i 
\epsilon_{\mu\nu\rho\sigma}k_{1}^{\rho}
p'^{\sigma}\nonumber \\&+& g_{\mu\nu}\frac{m_{B_{s}}^{2}-m_{\phi}^{2}}{2}  \}
F_{1}(0),
\label{A5}
\end{eqnarray}
where $\epsilon_{1 \mu}$, $\epsilon_{\nu}^{\phi}$ are the polarization vectors
and $k_{1}$, $p'$ are the momenta of the photon and
$\phi$ meson, respectively.
We used $G(k_1^{2}=0)=F_{1}(k_1^{2}=0)$ \cite{alibraun}.
Note, that the form factors introduced above are in general functions of two 
variables $k_1^2$ and $p'^2$. We abbreviated here 
$F_1(k_1^2) \equiv F_1(k_1^2, p'^2=m_{\phi}^2)$ and
took $F_1(0)=0.24 \pm 0.02$ from \cite{gudi}.

Now we want to compare our result for ${\cal{A}}(B_{s}\rightarrow\phi\gamma)$
eq.~(\ref{A5})
with the same amplitude calculated by the method worked out in
\cite{Ruckl}. This method is based on the new effective quark-gluon operator
obtained by the interaction of the virtual charm quark loop with soft gluons,
in contrast to a phenomenological description in terms of $\psi$ resonances 
converting into a photon, as we used. In this approach, the operator $O_{1}$
does not give any contribution to the matrix element of
$B_{s}\rightarrow\phi\gamma$ for an on-shell photon. 
The Fierz transformation of the
operator $O_{2}$ reads
\begin{eqnarray}
O_{2}=1/N_{C} \,O_{1}+1/2 \,\,O_{octet} \, \, ,
\label{O2exp}
\end{eqnarray}
where  
\begin{eqnarray}
O_{octet}=4 (\bar{c}\,
\gamma_{\mu}\frac{1-\gamma_{5}}{2}\frac{\lambda_{a}}{2}\, c)
(\bar{s}\,
\gamma_{\mu}\frac{1-\gamma_{5}}{2}\frac{\lambda_{a}}{2}\, b) \, \, ,
\label{octet}
\end{eqnarray}
and $\lambda^{a}/2$ are the $SU(3)$ colour generators.
Then the only contribution comes from the colour octet part $O_{octet}$.
Using the operator $O_{octet}$ as a vertex of the virtual charm quark loop,
which emits a real photon, and taking into account the c-quark-soft gluon 
interaction, a new effective operator is obtained. The matrix element 
of this operator between $B_{s}$ and $\phi$ meson states gives the long
distance amplitude of $B_{s}\rightarrow \phi\gamma$ decay 
due to the $O_{1,\,2}$ operators and it is written as 
(see \cite{Ruckl} for details; there the amplitude for the decay 
$B \to K^{\ast} \gamma$ is given)
\begin{eqnarray}
{\cal{A'}}(B_{s}\rightarrow\phi\gamma)=
2 C' \epsilon_{1}^{\mu} \epsilon^{\phi\, \nu}
\{ i \epsilon_{\mu\nu\rho\sigma} k_{1}^{\rho}p'^{\sigma}\, L+
\frac{m_{B_{s}}^{2}-m_{\phi}^{2}}{2} g_{\mu\nu}\, \tilde{L} \}\,,
\label{Aruckl}
\end{eqnarray}
where $C'=\frac{e G_{F}\lambda_{t}}{8 \sqrt{2}\pi^{2}}\frac{C_{2}(\mu)}{9
m_{c}^{2}}$ . 
The form factors $L$ and $\tilde{L}$ are calculated 
using QCD sum rules \cite{Ruckl},
\begin{eqnarray}
L&=& \frac{m_{b}}{m_{\phi} m_{B_{s}}^{2} f_{B_{s}} f_{\phi}} 
\,exp\,(\frac{m_{B_{s}}^{2}}{M^{2}}+\frac{m_{\phi}^{2}}{M'^{2}})
\nonumber \\  
&.& 
\{ \frac{m_{b}}{48}
\{ \frac{\alpha_{s}}{\pi} <G^2> \int_{m_{b}^{2}/M^{2}}^{\infty} ds\,\, 
e^{-s}\,
[ \frac{m_{b}^{2}}{s}-\frac{M^{4}}{M^{2}+M'^{2}}
(1-\frac{m_{b}^{2}}{s M^{2}})(1+\frac{M'^{2}}{s M^{2}})]\nonumber \\
&-& [\frac{m_{0}^{2} <\bar{s}s> m_{b}^{2}}{12}-\frac {4 \pi\alpha_{s}
<\bar{s}s>^{2} m_{b}}{27} (1+\frac{m_{b}^{2}}{M'^{2}})]
exp(-\frac{m_{b}^{2}}{M^{2}}) \} \nonumber\,\, , \\
\tilde{L}&=& \frac{m_{b}}{m_{\phi} m_{B_{s}}^{2} f_{B_{s}} f_{\phi}} 
exp(\frac{m_{B_{s}}^{2}}{M^{2}}+\frac{m_{\phi}^{2}}{M'^{2}}) 
\nonumber \\  
&.&
\{ \frac{m_{b}}{48}
\{ \frac{\alpha_{s}}{\pi}<G^2> \int_{m_{b}^{2}/M^{2}}^{\infty} ds\,\, 
e^{-s} \,
[ \frac{m_{b}^{2}}{s}+\frac{M^{4}}{M^{2}+M'^{2}}(1-\frac{m_{b}^{2}}{s
M^{2}})(1+\frac{M'^{2}}{s M^{2}})]\nonumber \\
&-& [\frac{m_{0}^{2} <\bar{s}s> m_{b}^{2}}{12}-\frac {16 \pi\alpha_{s}
<\bar{s}s>^{2} m_{b}}{27} (1+\frac{m_{b}^{2}}{M'^{2}})]
exp(-\frac{m_{b}^{2}}{M^{2}}) \}. 
\label{LLt}
\end{eqnarray}
The Borel parameters $M$ and $M'$ are varied to find the
stability region for $L$ and $\tilde{L}$.
We used the following input parameters in the sum rules:
$m_{0}^{2}=0.8 \,\, GeV^{2}$, $<\bar{s} s>=-0.011\,\, GeV^{3}$,
$\frac{\alpha_{s}}{\pi} <G^{2}>=0.012\,\, GeV^{4}$, 
$f_{\phi}=0.23\,\, GeV$, 
$f_{B_{s}}=0.2\,\, GeV$, $m_{\phi}=1.019\,\, GeV$ and 
$m_{B_{s}}=5.369\,\, GeV$.

The stability region is reached for $6\, GeV^2 \leq M^{2}\leq 9\, GeV^2$ 
and $2\, GeV^2\leq M'^{2}\leq 4\, GeV^2$ and we get 
\begin{eqnarray}
L&=&(0.30 \pm 0.05)\,\, GeV^3 \nonumber \, \, , \\
\tilde{L}&=&(0.35\pm 0.05)\,\, GeV^3   \, \, .
\label{llt}
\end{eqnarray}
Writing the amplitude for $B_{s}\rightarrow\phi\gamma$ as
\begin{eqnarray}
{\cal{A^{(')}}}(B_{s}\rightarrow\phi\gamma)=
\epsilon_{1}^{\mu} \epsilon^{\phi\, \nu}
( i \epsilon_{\mu\nu\rho\sigma} k_{1}^{\rho} p'^{\sigma} \, A^{-(')}+
g_{\mu \nu} \, A^{+(')} ) \,,
\label{ampgen}
\end{eqnarray}
and using eq.~(\ref{A5}),~(\ref{Aruckl}) and (\ref{llt}), we can compare the 
coefficients obtained by two different methods and get
\begin{eqnarray}
\frac{|A^{-}-A'^{-}|}
{A^{-}}&=& 10 \% \, \, \nonumber ,\\
\frac{|A^{+}-A'^{+}|}
{A^{+}}&=& 5 \% \, \, .
\label{ratio}
\end{eqnarray}
This means, that the amplitudes agree within $10 \%$.

In our approach, the structure for the transition $b\rightarrow s
\psi^{T}\rightarrow s \gamma$ is proportional to
$\sigma_{\mu \alpha}\,\frac{1+\gamma_{5}}{2} k_1^{\alpha}$ 
(see eq.~\ref{A2}), since the
longitudinal part of the $\psi$ meson is disregarded to convert into 
a photon. Further, the form factors $F_{1}(k_{1}^{2})$ 
and  $G(k_{1}^{2})$ in eq.~(\ref{alibraun}) are related for a real photon
($k_{1}^{2}=0)$,
$F_{1}(0)=G(0)$. Therefore, in the amplitude ${\cal{A}}(B_s \to \phi \gamma)$
appears only one form
factor, which is $F_{1}(0)$ in eq.~(\ref{A5}).
However, the form factors $L$ and $\tilde{L}$ in 
${\cal{A}}'(B_s \to \phi \gamma)$ given in eq.~(\ref{Aruckl})
are not related. 
They are calculated separately 
using QCD sum rules and
this causes the difference between the ratios in eq.~(\ref{ratio}).
In spite of the fact that
the amplitudes $A^{\pm}$ and $A'^{\pm}$ are different from each other,
they coincide within the given approximation and theoretical uncertainties
lying in both methods.

We can now present the amplitude for $B_{s}\rightarrow\gamma\gamma$ due to the
chain reaction 
$B_{s}\rightarrow\phi\psi\rightarrow\phi\gamma\rightarrow\gamma\gamma$
using the intermediate propagator at zero momentum transfer
and the $\phi\rightarrow\gamma$ conversion vertex from the VMD model,
\begin{eqnarray}
<0|J_{\mu\,\, el}|\phi(p',\epsilon^{\phi})>=e Q_{s} f_{\phi}(0) m_{\phi}
\epsilon_{\mu}^{\phi},
\label{Jel}
\end{eqnarray}
where the polarization vector $\epsilon_{\mu}^{\phi}$ is treated as
transversal.
To apply the VMD mechanism to the amplitude eq.~(\ref{A5}), we have to 
know the form factor at $F_1(k_1^2=0,p'^2=0)$. We took the extrapolated value
$\bar{F}_{1}(0)\equiv F_1(0,0)=0.16\pm 0.02 $
from \cite{gudi}. 
Then the amplitude can be written with $p' \to k_2$, 
$ \epsilon^{\phi} \to \epsilon_2$ as
\begin{eqnarray}
{\cal{A}}
(B_{s}\rightarrow\phi\psi\rightarrow\phi\gamma\rightarrow\gamma\gamma)=
\epsilon_{1}^{\mu}(k_{1})\epsilon_{2}^{\nu}(k_{2}) [ g_{\mu\nu}
A^{+}_{\bar{LD}_{O_{2}}}+i \epsilon_{\mu\nu\alpha\beta}k_{1}^{\alpha} k_{2}^{\beta}
A^{-}_{\bar{LD}_{O_{2}}} ]
\label{A6}
\end{eqnarray}
with the CP-even $A^{+}$ and CP-odd $A^{-}$ parts
\begin{eqnarray}
A^{+}_{\bar{LD}_{O_{2}}}&=&4 \chi \frac{C}{m_{b}} \bar{F}_{1}(0)
\sum_{i} f^{2}_{\psi_{i}}(0) e Q_{c}
\frac{m_{B_{s}}^{2}-m_{\phi}^{2}}{2} \, \, , \nonumber \\ 
A^{-}_{\bar{LD}_{O_{2}}}&=&4 \chi \frac{C}{m_{b}} \bar{F}_{1}(0)
\sum_{i} f^{2}_{\psi_{i}}(0) e Q_{c} \, \, ,
\label{Apm}
\end{eqnarray}
where $C$ is defined in eq.~(\ref{const1}) and 
the conversion factor $\chi$ is defined as 
$\chi=-e Q_{s} \frac{f_{\phi}(0)}{m_{\phi}}$.
Here $f_{\phi}(0)=0.18 \, \, GeV$ 
\cite{terasaki} and $Q_{s}=-1/3$.
The extra factor 2 comes from the addition of the diagram with interchanged 
photons.

The decay width for $B_{s}\rightarrow\gamma\gamma$ is obtained by adding the
short distance amplitudes, the $O_{7}$ type LD effects 
\cite{gudi} and the LD effects due to the 
$B_{s}\rightarrow\phi\psi\rightarrow\phi\gamma\rightarrow\gamma\gamma$
transition, resulting in
\begin{eqnarray}
\Gamma(B_{s}\rightarrow\gamma\gamma)
=\frac{1}{32\pi m_{B_{s}}}(
4|A^{+}+A^{+}_{LD_{O_{7}}}+A^{+}_{\bar{LD}_{O_{2}}}|^{2} \nonumber \\
+\frac{m^{4}_{B_{s}}}{2} |A^{-}+A^{-}_{LD_{O_{7}}}+A^{-}_{\bar{LD}_{O_{2}}}|^{2}) \, .
\label{gamma}
\end{eqnarray}
\section{Numerical estimates and discussion}
We presented a VMD model based calculation of the  LD contribution
to CP-even $A^{+}$ and CP-odd $A^{-}$ decay amplitudes for
$B_{s}\rightarrow\gamma\gamma$ decay due to the inclusive process
$b\rightarrow s\psi$. The conversions to photons from both the $\psi_{i}$ 
resonances
and the $\phi$ meson lead to two suppressions and make the amplitudes
in eq.~(\ref{Apm}) smaller compared to the ones from the LD effect of the 
operator $O_{7}$ \cite{gudi}.

We estimated the ratio
\begin{eqnarray}
\rho=|\frac{A^{+(-)}_{\bar{LD}_{O_{2}}}(
B_{s}\rightarrow\phi\psi\rightarrow\phi\gamma\rightarrow\gamma\gamma)}
{A^{+(-)}_{LD_{O_{7}}}(B_{s}\rightarrow\phi\gamma\rightarrow\gamma\gamma)}|
=4 \pi^{2} Q_c \frac{a_2^{eff}}{|C_7^{eff}(\mu)|}
\sum_i \frac{f_{\psi_{i}}^2(0)}{m_b^2}
\label{rho}
\end{eqnarray}
and found 
\begin{eqnarray}
2\% \leq \rho \leq 4 \%
\label{num}
\end{eqnarray}
while varying $\frac{m_b}{2} \leq \mu \leq 2 m_b$ and allowing
$a_2^{eff}$ to have a theoretical error of $25 \%$ as stated in 
\cite{neubert}. 
The analytical expression of the "effective" coefficient $C_7^{eff}(\mu)$
of the operator $O_7$ can be found in \cite{effham}.

As a conclusion, we investigated the LD-contribution to the
$B_{s}\rightarrow\gamma\gamma$ decay resulting from intermediate
$\psi_{i}$ production
and compared our result with the one obtained by the
interaction of the virtual charm loop with soft gluons \cite{Ruckl}.
We see that both amplitudes are in good agreement within the errors of the 
calculation. 
The new LD contribution resulting from the four-quark operators $O_1$ and
$O_2$ is smaller compared to the one coming from the
$B_{s}\rightarrow\phi\gamma\rightarrow\gamma\gamma$ chain decay \cite{gudi}
and effects our old estimate \cite{gudi} for the 
branching ratio
${\cal{B}}(B_{s}\rightarrow\gamma\gamma)_{SD+LD_{O_{7}}} $
not more than $1\%$.


\begin{thebibliography}{1}
\bibitem{effham}
A. J. Buras et al., Nucl. Phys. B {\bf 424} (1994) 374. \\
        A. Ali and C. Greub, Z. Phys. C {\bf 49} (1991) 431. 
\bibitem{yaosimmaaliev}
        G.-L. Lin, J. Liu and Y.-P. Yao, Phys. Rev. Lett. Vol. {\bf 64} No. {\bf 13} (1990) 1498; 
        G.-L. Lin, J. Liu and Y.-P. Yao, Phys. Rev. D {\bf 42} (1990) 2314;
        H. Simma and D. Wyler, Nucl. Phys. B {\bf 344} (1990) 283;
        T. M. Aliev and G. Turan, Phys. Rev. D {\bf 48} (1993) 1176.
\bibitem{gudi}
        G. Hiller and E. O. Iltan, Phys. Lett. B {\bf 409} (1997) 344.

\bibitem{desphande}
        N. G. Desphande, Xiao-Gang He and J. Trampetic, Phys. Lett.
B {\bf 367} (1996) 362. 
\bibitem{alibraun} A. Ali and V. M. Braun, Phys. Lett. B {\bf 359} (1995) 223.
\bibitem{pakvasa}
        E. Golowich and Pakvasa, Phys. Rev. D {\bf 51} (1995) 1215.
\bibitem{Ruckl}
        A. Khodjamirian, R. R\"uckl, G. Stoll and D. Wyler, hep-ph/9702318.
\bibitem{BSW} M. Bauer, B. Stech and M. Wirbel, Z. Phys C {\bf 34} (1987) 103.
\bibitem{neubert} M. Neubert and B. Stech, hep-ph/9705292, to appear in
the Second Edition of Heavy Flavours, edited by A. J. Buras and M. Lindner 
(World Scientific, Singapore).
\bibitem{PDG} R. M. Barnett et al., Review of Particle Properties, Phys Rev D
{\bf 54} (1996) 1.
\bibitem{terasaki} K. Terasaki, Nuovo Cim. Vol. {\bf 66} A, No. {\bf 4} (1981) 475.
\end{thebibliography}
\end{document}